\documentclass[prd,twocolumn,showpacs,amsmath,amssymb]{revtex4}

\usepackage{amssymb}
\usepackage{mathrsfs}
\usepackage{txfonts}

\usepackage{graphicx}
\usepackage{dcolumn}
\usepackage{bm}

\begin{document}

\title{Consistency of the warm k-inflation}

\author{Zhi-Peng Peng}
\email{zhipeng@mail.bnu.edu.cn}
\affiliation{Department of Physics, Beijing Normal University, Beijing 100875, China}

\author{Jia-Ning Yu}
\email{yujn@mail.bnu.edu.cn}
\affiliation{Department of Physics, Beijing Normal University, Beijing 100875, China}

\author{Xiao-Min Zhang}
\email{zhangxm@mail.bnu.edu.cn}
\affiliation{School of Science, Qingdao University of Technology, Qingdao 266033, China}

\author{Jian-Yang Zhu}
\thanks{Corresponding author}
\email{zhujy@bnu.edu.cn}
\affiliation{Department of Physics, Beijing Normal University, Beijing 100875, China}

\date{\today}
\begin{abstract}
We extend the k-inflation which is a type of kinetically driven inflationary model under the standard inflationary scenario to a possible warm inflationary scenario. The dynamical equations of this warm k-inflation model are obtained. We rewrite the slow-roll parameters which are different from the usual potential driven inflationary models and perform a linear stability analysis to give the proper slow-roll conditions in the warm k-inflation. Two cases, a power-law kinetic function and an exponential kinetic function, are studied, when the dissipative coefficient $\Gamma=\Gamma_0$ and $\Gamma=\Gamma(\phi)$, respectively. A proper number of e-folds is obtained in both concrete cases of warm k-inflation. We find a constant dissipative coefficient ($\Gamma=\Gamma_0$) is not a workable choice for these two cases while the two cases with $\Gamma=\Gamma(\phi)$ are self-consistent warm inflationary models.
\end{abstract}
\maketitle

\section{\label{sec1}Introduction}
The inflationary paradigm \cite{Guth1981,Linde1982,Albrecht1982} gives an attractive explanation to resolve many of the puzzles (horizon, flatness and monopoles) of the standard cosmological model. It is generally considered that the inflation produced seeds which give rise to the large scale structure and the observed little anisotropy of cosmological microwave background (CMB) \cite{Planck2015} through the vacuum fluctuations. During the standard inflation, sometimes called cold inflation, the Universe undergoes a steep supercooling phase, for it is assumed that the scalar field which drives the inflation is isolated and the interaction between the inflaton field and other fields can be neglected. After the supercooling phase, the Universe needs a reheating epoch to become hot again and is filled with radiation required by the standard cosmology. Besides the standard inflation, there exists another type of inflation called the warm inflation proposed by Berera and Fang \cite{Berera1995}. Radiation is produced constantly through the interaction between the inflaton field and other subdominated boson or fermion fields during the warm inflation so that there is no reheating phase. And the density fluctuations originate mainly from the thermal fluctuations \cite{Berera1995,Hall2004} rather than the vacuum fluctuations. In the above scenario, $\mathcal{L}_{int}$ in the Lagrangian density of the scalar field describes the interaction of inflaton with all other fields. The friction term including $\Gamma\dot{\phi}$ in the equation of motion of the inflaton field phenomenologically describes the decay of the inflaton field into the thermal bath via the interaction Lagrangian $\mathcal{L}_{int}$ \cite{Berera1995,Berera1996}. And matters in the Universe can be generated by the decay of the inflaton field or the radiation field \cite{Taylor1997}. In warm inflation, many problems such as $\eta$-problem \cite{Dine1995} and the overlarge amplitude of the inflaton field \cite{Berera2005,Berera2009} suffered in standard inflation can be avoided. Furthermore, the slow-roll conditions are much more easily satisfied in warm inflationary scenario \cite{del2010, Zhang2013}.

Generally, the potential energy of the inflaton field  in the cold or warm inflation  is the dominating energy and drives the inflation during the inflationary epoch. However, a novel model called ``k-inflation'' provided by Mukhanov \cite{Mukhanov1999} only has a general class of non-standard (i.e. non-quadratic) $kinetic$-$energy$ terms for a scalar field $\phi$. And it turned out to drive an inflationary evolution of the same type as the usually considered potential driven inflation. The consideration of non-standard kinetic terms is appealed to the existence, in string theory, of higher-order corrections to the effective action of the scalar field $\phi$. Therefore, it provides a new mechanics for implementing inflation. In this paper we extend the k-inflation to a warm inflationary paradigm and obtain the dynamical equations {\cite{Ramos1998}}. Then we make use of a slow-roll approximation which neglects the highest order term  to simplify the dynamical equations of the system. We study strictly whether the slow-roll equations can describe the inflation exactly by a linear stability analysis used to obtain the slow-roll conditions for the system to remain close to the slow-roll solution for many Hubble times \cite{del2010,Moss2008,Mar2012}. The slow-roll conditions we obtained are similar to those in the cold inflation
rather than the usual standard inflationary models extended to warm inflation \cite{XiaoMin2014}. Then through two concrete examples, a power-law kinetic function and an exponential kinetic function, we find that a constant dissipative coefficient ($\Gamma=\Gamma_0$) is not a suitable choice while the two cases with $\Gamma=\Gamma(\phi)$ are self-consistent warm k-inflation models.

The paper is organized as follows. In Sec.\ref{sec2}, we introduce the basic dynamics of the warm k-inflation. Then in Sec.\ref{sec3}, we rewrite the new slow-roll parameters which are very different from those in the usual inflationary models. Then the linear stability analysis of the warm k-inflation is performed in Sec.\ref{sec4} and two concrete examples are discussed in Sec.\ref{sec5} and Sec.\ref{sec6} respectively. Finally, we draw the conclusions in Sec.\ref{sec7}.

\section{\label{sec2}dynamics of warm k-inflation}

In warm inflationary case, the Universe is constructed by multi-components;  thus, the total matter action can be written as
\begin{equation}\label{action}
  S=\int d^4x \sqrt{-g}  \left[ \mathcal{L}(X,\phi)+\mathcal{L}_R+\mathcal{L}_{int}\right],
\end{equation}
where  $\mathcal{L}(X,\phi)$ is the Lagrangian density of the inflaton field, $\mathcal{L}_R$ is the Lagrangian density of radiation field and $\mathcal{L}_{int}$ describes the interaction between the inflaton field and other fields. The Lagrangian density of inflaton field takes the same simple form as that in the Ref.\cite{Mukhanov1999},
\begin{equation}\label{field}
  \mathcal{L}(X,\phi)=K(\phi)X+X^2,
\end{equation}
where $K(\phi)$ is a function of the inflaton field $\phi$, called `` kinetic function'' and $X=\frac{1}{2} g^{\mu\nu}\partial_{\mu}\phi\partial_{\nu}\phi$. The Lagrangian density should satisfy the conditions: $\mathcal{L}_{X}\geq 0$ and $\mathcal{L}_{XX}\geq 0$ (where a subscript $X$ here means a partial derivative with respect to $X$ while the subscripts in $\mathcal{L}_{R}$ and $\mathcal{L}_{int}$ are just labels) to obey the null energy condition and the physical propagation of perturbations \cite{Bean2008,Franche2010}.

The inflaton field in a spatially flat Friedmann-Robertson-Walker (FRW) Universe is described by an effective fluid with the energy momentum tensor $T^{\mu}_{\phantom{\mu}\nu}=diag(\rho_{\phi},-p_{\phi},-p_{\phi},-p_{\phi})$, where the energy density $\rho_{\phi}$ and the pressure $p_{\phi}$ of the inflaton field are given by
\begin{equation}\label{density}
  \rho_{\phi}=K(\phi)X+3X^2,
\end{equation}
and
\begin{equation}\label{pressure}
  p_{\phi}=K(\phi)X+X^2,
\end{equation}
respectively. Besides, the four-velocity of the inflaton field is $U^{\mu}=\sigma\frac{\nabla^{\mu}\phi}{\sqrt{2X}}$, where $\sigma$ denotes the sign of $\dot{\phi}$.

The Friedmann equation is given by
\begin{equation}\label{Friedmann}
  H^2=\frac{1}{3M_{p}^2}\rho,
\end{equation}
where $M_{p}^2\equiv(8\pi G)^{-1}$ and $\rho$ is the total energy density of the Universe. As usual, we consider an homogeneous background scalar field $X=\frac{1}{2}\dot{\phi}^2$, then the equation of motion of the inflaton field can be obtained by the variation of the Lagrangian density of inflaton field (\ref{field})
\begin{equation}\label{EOM}
  (3\dot{\phi}^2+K)\ddot{\phi}+3H(\dot{\phi}^2+K)\dot{\phi}+\Gamma\dot{\phi}+\frac{1}{2}K_{\phi}\dot{\phi}^2=0,
\end{equation}
where $K_{\phi}$ is a derivative with respect to $\phi$, and on the basis of the thermal dissipation assumption the term $\Gamma \dot{\phi}$ we add like the canonical warm inflation \cite{Berera1995} phenomenologically describes the decay of the inflaton field $\phi$ via the interaction Lagrangian $\mathcal{L}_{int}$ during the inflationary phase. However, the dissipative coefficient $\Gamma$ here has different functional form from that in canonical warm inflation. Giving the specific functional form of $\Gamma$ need further research of the microphysical basis of the non-canonical warm inflation, which we will study later like the Ref.\cite{Moss2009}. In principle, just as the Refs. \cite{Berera1995,XiaoMin2014} said, the term $\Gamma \dot{\phi}$ may not be reasonable for denotes the energy transfer from inflaton field $\phi$ during far out of equilibrium conditions. However, it is a proper approximation for the energy dissipated by the inflaton field $\phi$ into a thermalized radiation bath. In addition, the evolution equation for the energy density $\rho_{\phi}$ of inflaton field is given by
\begin{equation}\label{phicon}
\dot{\rho}_{\phi}+3H(\rho_{\phi}+p_{\phi})=-\Gamma\dot{\phi}^2,
\end{equation}
where we have assumed that the interaction between non-canonical inflaton field and other fields has the same form as those in canonical case {\cite{Bastero2009}}. Then we find it is self-consistent because when we substitute Eqs.(\ref{density}) and (\ref{pressure}) into Eq.(\ref{phicon}), we can get the same equation of motion of the inflaton field as Eq.(\ref{EOM}).

Since during the inflation, the dominant components of the Universe are the inflaton field and the radiation, so the total energy density $\rho$ and pressure $p$ are written as
\begin{equation}\label{rho}
\rho=\rho_{\phi}+\rho_{\gamma}=K(\phi)X+3X^2+\rho_{\gamma},
\end{equation}
and
\begin{equation}\label{p}
p=p_{\phi}+p_{\gamma}=K(\phi)X+X^2+\frac{1}{3}\rho_{\gamma},
\end{equation}
where $\rho_{\gamma}$ is radiation density and $p_{\gamma}$ is radiation pressure .
Furthermore, according to the total energy-momentum conservation equation, $\dot{\rho}+3H(\rho+p)=0$ and Eq.(\ref{phicon}), we can get the equation of production of radiation
\begin{equation}\label{radiation}
\dot{\rho}_{\gamma}+4H\rho_{\gamma}=\Gamma \dot{\phi}^2.
\end{equation}
Generally, we consider radiation production is quasi stable \cite{Berera1995}, i.e. $\dot{\rho}_{\gamma}\ll4H\rho_{\gamma}$. From Eq.(\ref{radiation}) we obtain that the density of radiation becomes
\begin{equation}\label{approximateradiation}
\rho_{\gamma}=\kappa T^4\simeq\frac{3}{4}r\,\dot{\phi}^2,
\end{equation}
where $\kappa$ is the Stefan-Boltzmann constant, $T$ is the temperature of the thermal bath and the dissipative rate $r\equiv\frac{\Gamma}{3H}$.

From the Friedmann equation (\ref{Friedmann}) and the total energy-momentum conservation equation, we can get the master equation of the Universe
\begin{equation}\label{state}
\dot{\rho}=-\frac{\sqrt{3\rho}}{M_{p}}(\rho+p).
\end{equation}
In order to obtain a quasi-exponential inflation, the total energy density and the pressure satisfy approximately the relation, $p\simeq -\rho$, thus we can obtain
\begin{equation}\label{prho}
4X^2+2KX+\frac{4}{3}\rho_{\gamma}\simeq 0
\end{equation}
during the inflation. In terms of Eqs.(\ref{radiation}) and (\ref{prho}), we can get the attractive fixed point of Eq.(\ref{state}),
\begin{equation}\label{fixedpoint}
X_{f}=-\frac{K(\phi)+r}{2},
\end{equation}
where $K(\phi)<0$ since the pressure $p<0$ during the inflation. If $\Gamma=0$ i.e. $r=0$, the attractive fixed point becomes $X_{f}=-\frac{1}{2}K(\phi)$ which is consistent with that in the standard inflation \cite{Mukhanov1999}. Then from the Friedmann equation (\ref{Friedmann}) we can obtain
\begin{equation}
H_{f}=\frac{\sqrt{K(\phi)[K(\phi)+r]}}{2\sqrt{3}M_{p}}.
\end{equation}
The number of e-folds of inflation is given by
\begin{equation}\label{efold}
N=\int_{t_i}^{t_f}Hdt=\int_{\phi_{i}}^{\phi_{e}}\frac{H}{\dot{\phi}}d\phi\simeq\frac{\sigma}{2\sqrt{3}M_{p}}\int_{\phi_{i}}^{\phi_{e}}\sqrt{-K(\phi)}d\phi,
\end{equation}
where $\phi_{i}$ is the initial value of the inflaton field and $\phi_{e}$ is the final value of the inflaton field.

\section{\label{sec3}slow-roll parameters}

Inflationary solutions to the exact equations (\ref{Friedmann}), (\ref{EOM}) and (\ref{radiation}) are difficult to calculate, so a slow-roll approximation is often applied. The slow-roll approximation usually neglects the highest order terms in the exact equations thus we need some slow-roll parameters to explain the rationality of approximation for convenience. However, since the warm k-inflation is very different from the usual potential driven inflationary models, we need rewrite the slow-roll parameters in warm k-inflation.  For convenience, we define a new variable $u=\dot{\phi}$. Then Eqs.(\ref{Friedmann}), (\ref{EOM}) and (\ref{radiation}) can be rewritten as
\begin{eqnarray}\label{Friedmann1}
H^2=\frac{1}{3M_{p}^2}\left(\frac{3}{4}u^4+\frac{K}{2}u^2+\rho_{\gamma}\right),
\end{eqnarray}
\begin{eqnarray}\label{EOM1}
(3u^2+K)\dot{u}+3H(u^2+K+r)u+\frac{1}{2}K_{\phi}u^2=0,
\end{eqnarray}
\begin{eqnarray}\label{radiation1}
\dot{\rho}_{\gamma}+4H\rho_{\gamma}=\Gamma u^2.
\end{eqnarray}

Now we begin to rewrite the  slow-roll parameters according to the following requirements of the quasi exponential inflation.

(a) In order to obtain a quasi exponential inflation, the fractional change $\vert \frac{\dot{H}}{H}\vert \frac{1}{H}$ in $H$ during an expansion time $\frac{1}{H}$ must be much less than unity \cite{Weinberg}, i.e. $-\frac{\dot{H}}{H^2}\ll1$, where the `` dot '' is a derivative with respect to time. Then on the basis of Eqs.(\ref{approximateradiation}) and (\ref{Friedmann1}), we can find $-\frac{\dot{H}}{H^2}\simeq\frac{6}{3-K/(u^2+K+r)}\ll1$. Thus, we can get a necessary condition for the accelerated expansion in warm k-inflation,
\begin{equation}
\epsilon\equiv-\frac{3(u^2+K+r)}{K}\ll 1.
\end{equation}
This means that Eq.(\ref{Friedmann1}) can be rewritten nearly as
\begin{equation}\label{slowrolleq1}
H^2=\frac{1}{3M_{p}^2}\frac{3}{4}u^2\bigg(u^2+K+r-\frac{1}{3}K\bigg)\simeq\frac{1}{3M_{p}^2}\bigg(-\frac{1}{4}Ku^2\bigg).
\end{equation}
So, during the inflation, the kinetic term $-\frac{1}{4}Ku^2$ is the dominating energy, or $\rho_{\gamma}\ll\rho_{\phi}$.

(b) In order to get enough number of e-folds, the fractional change $\vert \frac{\ddot{H}}{\dot{H}}\vert \frac{1}{H}$ in $H$ during an expansion time $\frac{1}{H}$ must be much less than unity, i.e.
\begin{equation}\label{Htt}
\bigg\vert \frac{\ddot{H}}{H\dot{H}} \bigg\vert=\bigg \vert 2\frac{\dot{u}}{Hu}+\frac{(u^2+K+r)^{\mathbf{\cdot}}}{H(u^2+K+r)}\bigg \vert \ll1.
\end{equation}
Usually, we will assume the fractional change $\vert \frac{\ddot{\phi}}{\dot{\phi}}\vert \frac{1}{H}$ in $H$ during an expansion time $\frac{1}{H}$ should be much less than unity {\cite{Weinberg}}, i.e. $ \big\vert \frac{\dot{u}}{Hu}\big\vert\ll 1$. Thus, the sufficient condition for the establishment of Eq.(\ref{Htt}) is that $\big\vert \frac{(u^2+K+r)^{\mathbf{\cdot}}}{H(u^2+K+r)}\big\vert \ll1$. Then $\big\vert \frac{\dot{u}}{Hu} \big\vert\ll 1$ means $u$ varies slowly during the inflation, thus Eq.(\ref{EOM1}) can be rewritten nearly as
\begin{equation}\label{slowrolleq2}
  3H(u^2+K+r)u+\frac{1}{2}K_{\phi}u^2=0.
\end{equation}
According to the above equation, the first slow-roll parameter $\epsilon$ in warm k-inflation can also be rewritten as
\begin{equation}
\epsilon=-\frac{3(u^2+K+r)}{K}\simeq\frac{K_{\phi}u}{H K}.
\end{equation}
In addition, on the basis of Eq.(\ref{slowrolleq2}), we can obtain
\begin{equation}\label{slow2}
\bigg\vert \frac{(u^2+K+r)^{\mathbf{\cdot}}}{H(u^2+K+r)}\bigg \vert=\bigg\vert \frac{K_{\phi\phi}u}{HK_{\phi}}+\frac{\dot{u}}{Hu} +\epsilon \bigg\vert \ll1,
\end{equation}
because $\epsilon\ll1$ and $\big\vert \frac{\dot{u}}{Hu} \big\vert\ll 1$, the sufficient condition for the establishment of Eq.(\ref{slow2}) is that ,
\begin{equation}\label{slow3}
\vert \eta \vert \equiv \bigg\vert\frac{K_{\phi\phi}u}{HK_{\phi}}  \bigg\vert \ll1,
\end{equation}
where $K_{\phi\phi}$ is a two-order derivative with respect to the inflaton field $\phi$. Thus we have get the second slow-roll parameter $\eta$ in warm k-inflation. $\vert \eta \vert \ll1$ implies that $K_{\phi}$ varies slowly during the inflation.

(c) In Sec.{\ref{sec2}}, we assume that the radiation production is quasi stable, which means
\begin{equation}\label{slow1}
\bigg\vert \frac{\dot{\rho}_{\gamma}}{4H\rho_{\gamma}} \bigg\vert= \bigg\vert \frac{\dot{r}}{4Hr}+\frac{\dot{u}}{2Hu} \bigg\vert \ll1.
\end{equation}
Because $\big\vert \frac{\dot{u}}{Hu} \big\vert \ll1 $ , the sufficient condition for the establishment of Eq.(\ref{slow1}) is that $\big\vert \frac{\dot{r}}{Hr} \big\vert \ll1 $. And $\big\vert \frac{\dot{r}}{Hr} \big\vert \ll1$ also means that
\begin{equation} \label{slow4}
\bigg\vert \frac{\dot{r}}{Hr} \bigg\vert = \bigg\vert \frac{\Gamma_{\phi}u}{H\Gamma}+ \frac{\dot{\rho}_{\gamma}}{4H\rho_{\gamma}}c+\epsilon \bigg\vert \ll1,
\end{equation}
where $c\equiv\frac{T\Gamma_{T}}{\Gamma}$ is a finite quantity and $\Gamma_{T}$ is a partial derivative with respect to the temperature $T$. Because $\big\vert \frac{\dot{\rho}_{\gamma}}{4H\rho_{\gamma}} \big\vert \ll1 $ and $\epsilon\ll1$, the sufficient condition for the establishment of Eq.(\ref{slow4}) is that $\big\vert \frac{\Gamma_{\phi}u}{H\Gamma} \big\vert \ll1$. Thus we can obtain the third  slow-roll parameter in warm k-inflation,
\begin{equation}
  \vert b \vert \equiv \bigg\vert \frac{\Gamma_{\phi}u}{H\Gamma} \bigg\vert \ll1,
\end{equation}
where $\Gamma_{\phi}$ is a derivative with respect to the inflaton field $\phi$. Then $\vert b \vert \ll1$ means that the dissipative coefficient $\Gamma$ varies slowly with the inflaton field.

Now in order to obtain enough number of e-folds of the quasi exponential inflation, we have obtain three independent slow-roll parameters in our model as follows
\begin{eqnarray}
\epsilon=\frac{K_{\phi}u}{HK},\quad \eta=\frac{K_{\phi\phi}u}{HK_{\phi}},\quad b=\frac{\Gamma_{\phi}u}{H\Gamma},
\end{eqnarray}
which should be much less than unity. In addition, we define a finite quantity, $c$ , which is the same as the usual warm inflation \cite{Bastero2009} to describe the change of dissipative term $\Gamma$ with the temperature. And we find the constraints on the slow-roll parameters are equivalent to the usual slow-roll approximation  which result in the slow-roll equations (\ref{approximateradiation}), (\ref{slowrolleq1}) and (\ref{slowrolleq2}). Furthermore, the slow-roll approximation implies that the energy is kinetic energy dominated during the inflation which means that the warm k-inflation is a type of kinetically driven inflationary model.

\section{\label{sec4}stability analysis}

We have obtained slow-roll parameters in warm k-inflation which is very different from those in usual potential driven inflationary models, then we should show that under what conditions can the slow-roll equations (\ref{approximateradiation}), (\ref{slowrolleq1}) and (\ref{slowrolleq2}) describe the system well. Thus we will perform a linear stability analysis to get the sufficient conditions for the system to remain close to the slow-roll solutions for many Hubble times, i.e. the slow-roll solutions should be an attractor for the dynamical system. For the stability analysis, we use $\phi_{0}$, $u_{0}$ and $\rho_{\gamma_{0}}$ to express the slow-roll solutions which satisfy slow-roll equations below:
\begin{equation}\label{sleq1}
3H_{0}(u_{0}^2+K+r)u_0+\frac{1}{2}K_{\phi}u_{0}^2=0,
\end{equation}
\begin{equation}\label{sleq2}
4H_{0}\rho_{\gamma_{0}}-\Gamma u_{0}^2=0,
\end{equation}
\begin{equation}\label{sleq3}
H_{0}^2=\frac{1}{3M_{p}^2}\bigg(-\frac{1}{4}Ku_{0}^2\bigg).
\end{equation}
The exact solutions $\phi$, $u$ and $\rho_{\gamma}$ can be expanded around the slow-roll solutions: $\phi=\phi_{0}+\delta\phi$, $u=u_{0}+\delta u$ and $\rho_{\gamma}=\rho_{\gamma_{0}}+\delta\rho_{\gamma}$. The perturbations $\delta\phi$, $\delta u$ and $\delta\rho_{\gamma}$ are much smaller than the background ones, i.e. $\delta\phi\ll\phi_{0}$, $\delta u \ll u_{0}$ and $\delta\rho_{\gamma}\ll\rho_{\gamma_{0}}$. Besides, in the following stability analysis we will make use of the former slow-roll parameters frequently: $\epsilon=\frac{K_{\phi}u}{HK}$, $\eta=\frac{K_{\phi\phi}u}{HK_{\phi}}$ and $b=\frac{\Gamma_{\phi}u}{H\Gamma}$.

Varying the Friedmann equation (\ref{Friedmann1}), we can obtain
\begin{equation}
\delta H=\frac{1}{6M_{p}^2H_{0}}\Bigg[(3u_{0}^2+K)u_{0}\delta u+\frac{K_{\phi}}{2}u_{0}^2\delta\phi+\delta\rho_{\gamma}\Bigg],
\end{equation}
and varying the dissipative coefficient $\Gamma$, we get
\begin{equation}
\delta\Gamma=\Gamma_{\phi}\delta\phi+\frac{T\Gamma_{T}}{4\rho_{\gamma_{0}}}\delta\rho_{\gamma}.
\end{equation}
Then we can get the variations of $\delta \rho_{\gamma}$ and $\delta r$ in the similar method. These equations will be applied later.

Taking the variation of Eqs.(\ref{EOM1}) and (\ref{radiation1}), we can obtain
\begin{equation}
{\left( \begin{array}{ccc}
\dot{\delta\phi} \\
\dot{\delta u}  \\
\dot{\delta\rho_{\gamma}}\\
\end{array} \right)}=E\cdot{\left( \begin{array}{ccc}
\delta\phi \\
\delta u  \\
\delta\rho_{\gamma}\\
\end{array} \right)}-F.
\end{equation}
The matrices $E$ and $F$ are expressed as
\begin{eqnarray}
E=
  {\left(
    \begin{array}{ccc}
      0 & 1 & 0 \\
      A & \lambda_{1} & B\\
      C & D & \lambda_{2} \\
    \end{array}
  \right)}\, ,\quad F={\left(
                 \begin{array}{c}
                   0\\
                   \dot u_0\\
                   \dot \rho_{\gamma_{0}}\\
                 \end{array}
               \right)}.
\end{eqnarray}
Then the matrix elements of $E$ can be worked out
\begin{eqnarray}\label{A}
  A=\frac{3H_{0}^2}{\mathcal{L}_{X}}c_{s}^2\Bigg[\mathcal{L}_{X}\epsilon+(\mathcal{L}_{X}+r)\eta+6(\mathcal{L}_{X}+r)-rb\Bigg],
\end{eqnarray}
\begin{eqnarray}\label{B}
B=-\frac{H_0}{\mathcal{L}_{X}u_0}c_{s}^2\Bigg[\frac{\mathcal{L}_{X}}{\mathcal{L}_{X}+r}\epsilon+c\Bigg],
\end{eqnarray}
\begin{eqnarray}\label{C}
C=3H_0u_0r(\epsilon+b),
\end{eqnarray}
\begin{eqnarray}\label{D}
D=H_0u_0\Bigg(6r-\frac{r\mathcal{L}_{X}c_{s}^{-2}}{\mathcal{L}_{X}+r}\epsilon\Bigg),
\end{eqnarray}
\begin{eqnarray}\label{lambda1}
\lambda_{1}=-H_0\frac{\mathcal{L}_{X}}{\mathcal{L}_{X}+r}\epsilon-3H_0\Bigg[1-\Bigg(2+\frac{r}{\mathcal{L}_{X}}\Bigg)c_{s}^2\Bigg],
\end{eqnarray}
\begin{eqnarray}\label{lambda2}
\lambda_{2}=-H_0(4-c)-H_0\frac{r}{\mathcal{L}_{X}+r}\epsilon,
\end{eqnarray}
where $\mathcal{L}_{X}=u_{0}^2+K$, and $c_{s}^2=\frac{u_{0}^2+K}{3u_{0}^2+K}$ which is the perturbation propagation speed of the inflaton field \cite{Mukhanov}.

The slow-roll solution can be an attractor for the warm inflationary dynamics system only when the eigenvalues of the matrix $E$ are negative or possibly positive but of order $\mathcal{O}(\epsilon)$ and the``forcing term'' $F$ is small enough, i.e. $\big\vert \frac{\dot{u}_{0}}{H_0u_0} \big\vert \ll1$ and $\big\vert \frac{\dot{\rho}_{\gamma_0}}{H\rho_{\gamma_0}}\big\vert \ll1$.

Now we begin to study the forcing term $F$ firstly. Taking the time derivative of the slow-roll equations (\ref{sleq1}) and (\ref{sleq2}), we can obtain
\begin{eqnarray}\label{force1}
\frac{\dot{u}_0}{H_0u_0}&=&\frac{1}{\Delta}\Bigg[\bigg(4c_{s}^2-cc_{s}^2-\frac{crc_{s}^2}{\mathcal{L}_{X}}\bigg)\epsilon
+\frac{(\mathcal{L}_{X}+r)c_{s}^2}{\mathcal{L}_{X}}\bigg(4-c\bigg)\eta \nonumber\\
&-&4\frac{rc_{s}^2}{\mathcal{L}_{X}}b-6\frac{(\mathcal{L}_{X}+r)c_{s}^2}{\mathcal{L}_{X}}\bigg(c-4\bigg)\Bigg],
\end{eqnarray}
\begin{eqnarray}\label{force2}
\frac{\dot{\rho}_{\gamma_0}}{H_0\rho_{\gamma_0}}&=&\frac{8}{\Delta}\Bigg[\bigg(\frac{1}{2}-\frac{rc_{s}^2}{2\mathcal{L}_{X}}\bigg)\epsilon
+ \frac{(\mathcal{L}_{X}+r)c_{s}^2}{\mathcal{L}_{X}}\eta \nonumber\\
&-&\bigg(\frac{rc_{s}^2}{\mathcal{L}_{X}+r}+\frac{3rc_{s}^2}{2\mathcal{L}_{X}}-\frac{1}{2}\bigg)b+6\frac{(\mathcal{L}_{X}+r)c_{s}^2}{\mathcal{L}_{X}}\Bigg],
\end{eqnarray}
where $\Delta=(4-c)(1-2c_{s}^2)+(c-2)\frac{rc_{s}^2}{\mathcal{L}_{X}}$. Because $\big\vert \frac{\dot{u}_{0}}{H_0u_0} \big\vert \ll1$ and $\big\vert \frac{\dot{\rho}_{\gamma_0}}{H\rho_{\gamma_0}}\big\vert \ll1$, we take the absolute value of each term in Eqs.(\ref{force1}) and (\ref{force2}) to be individually much less $1$ to get the sufficient conditions of satisfying above requirements,
\begin{eqnarray}
\epsilon\ll1,\quad \vert\eta\vert\ll\frac{\mathcal{L}_{X}}{(\mathcal{L}_{X}+r)c_{s}^2}, \quad \vert b\vert \ll1,\quad
\frac{rc_{s}^2}{\mathcal{L}_{X}}\ll1-c_{s}^2.
\end{eqnarray}
From the last condition, we can find $\frac{\mathcal{L}_{X}}{(\mathcal{L}_{X}+r)c_{s}^2}<1$, which means $|\eta|\ll1$.

Then now we begin to study the matrix $E$ to obtain more constraints. On the basis of the above constraints, the elements $A$ and $C$ are much smaller than other elements of the matrix $E$. Thus from the eigenvalue equation of $E$
\begin{eqnarray}
det(\lambda I-E)&=&
  {\Bigg|
    \begin{array}{ccc}
      \lambda & -1 & 0 \\
      -A & \lambda-\lambda_{1} & B\\
      -C & -D & \lambda-\lambda_{2} \\
    \end{array}
  \Bigg|} \nonumber\\
  &=&\lambda(\lambda-\lambda_{1})(\lambda-\lambda_{2})-BD\lambda-A(\lambda-\lambda_{2})-BC\nonumber\\
  &=&0,
\end{eqnarray}
we can get a small eigenvalue $\lambda\simeq\frac{BC-A\lambda_{2}}{\lambda_{1}\lambda_{2}-BD-A}\ll\lambda_{1},\lambda_{2}$. The other two eigenvalues satisfy the equation: $\lambda^2-(\lambda_1+\lambda_2)\lambda+\lambda_1\lambda_2-BD=0$. When $\lambda_1+\lambda_2<0 $ and $\lambda_1\lambda_2-BD>0$, the two eigenvalues are both negative. Thus we take each term in Eqs.(\ref{lambda1}) and (\ref{lambda2}) to be individually smaller than $0$ so that we obtain the sufficient conditions of satisfying $\lambda_1<0$ and $\lambda_2<0$,
\begin{eqnarray}
\vert c \vert<4,\  \frac{rc_{s}^2}{\mathcal{L}_{X}}\ll1-2c_{s}^2.
\end{eqnarray}
To sum up, we have obtained the sufficient conditions or slow-roll conditions for the dynamic system to remain close to the slow-roll solutions for many Hubble times as follows,
\begin{eqnarray}
\epsilon\ll1,\ \vert \eta \vert \ll\frac{\mathcal{L}_{X}}{(\mathcal{L}_{X}+r)c_{s}^2},\vert b \vert \ll1,\ \vert c \vert <4, \frac{rc_{s}^2}{\mathcal{L}_{X}}\ll1-2c_{s}^2
\end{eqnarray}
where the first two conditions mean the slow-roll solutions are quasi exponential solutions, $ \vert b \vert \ll1$ means that $\Gamma$ is varying slowly together with the inflaton field. The constraint on parameter $c$ is the same as that of usual warm inflation models \cite{Bastero2009}. This implies that the dependence of the dissipative coefficient should be within in the range of $\Gamma\propto(T^{-4},T^4)$. The stability analysis of the warm k-inflation is applicable to the general dissipative coefficient $\Gamma$ which is the function of the infaton field $\phi$ and the temperature $T$.

Finally, we can  summarize the former constraints on the kinetic function $K(\phi)$ for the analysis of specific models later. In Sec.{\ref{sec2}}, we have known that $\mathcal{L}_{X}\geq 0$, $\mathcal{L}_{XX}\geq 0$, which imply $|K|\leq u^2$. And the pressure $p$ should be negative during the inflation, thus we get $|K|>\frac{1}{2}(u^2+r)$. Besides, the attractive fixed point in Sec.{\ref{sec2}} should be positive thus we can deduce that $|K|>r$, and by stability analysis we can get $\frac{rc_{s}^2}{\mathcal{L}_{X}}\ll1-2c_{s}^2$ which means $|K|\gg r-u^2$.  In all, we find the range of $K(\phi)$ is $\frac{1}{2}(u^2+r)<|K|\leq u^2$. Together with $3(u^2+K+r)\ll|K|$, we can know $u^2$ should be the same order of the amplitude as $|K|$ while $r$ should be smaller than both $u^2$ and $|K|$.

\section{\label{sec5}power-law form of $K(\phi)$}

In this section , we consider a kind of specific Lagrangian density to test the reliability of the warm k-inflation. We make the kinetic function $K(\phi)$ take the power-law form:
\begin{equation}\label{model1}
  K(\phi)=-K_{0}^2\phi^{2n},
\end{equation}
where $K_{0}$ is a positive constant and $n$ is a positive integer. This kind form of $K(\phi)$ has been showed that it can result in kinetically driven inflation in the Ref.\cite{Mukhanov1999}. And the dissipative coefficient $\Gamma$ take two kinds of forms $\Gamma=\Gamma_{0}=constant$ and $\Gamma=\Gamma(\phi)$ respectively in the following. Here unlike the refs.{\cite{Ramos2011, Rosa2013}}, for simplicity we assume the dissipative coefficient $\Gamma$ is independent of the temperature $T$ in our specific models as follows. However, the stability analysis in Sec.\ref{sec4} is truly applicable to the general dissipative coefficient $\Gamma(\phi,T)$.  Then we begin to test whether concrete warm k-inflation models can get enough number of e-folds of quasi exponential expansion and exit to reach the radiation dominated phase smoothly.

\subsection{$\Gamma=\Gamma_{0}$ case.}

When $\Gamma=\Gamma_{0}=constant$ and $K(\phi)=-K_{0}^2\phi^{2n}$, we can work out the Hubble parameter, dissipative rate and radiation density during the slow-roll inflationary phase:
\begin{eqnarray}
  H&=&\frac{K_{0}\phi^{n}|u|}{2\sqrt{3}M_{p}},\label{H}\\
  r&=&\frac{2\Gamma_{0}M_{p}}{\sqrt{3}K_0\phi^{n}|u|},\label{r}\\
  \rho_{\gamma}&=&\frac{\sqrt{3}\Gamma_{0}M_{p}|u|}{2K_{0}\phi^{n}}.
\end{eqnarray}
In Sec.\ref{sec5}, we have known $u^2$ has the same order of the amplitude as $|K(\phi)|$. Thus we can estimate that $u\propto\phi^{n}$, then $r\propto\phi^{-2n}$. Moreover, the slow-roll parameters can be given by
\begin{eqnarray}
  \epsilon&=&\frac{4\sqrt{3}nM_{p}}{K_{0}\phi^{n+1}},\label{epsilon1}\\
  \eta&=&\frac{2\sqrt{3}(2n-1)M_{p}}{K_{0}\phi^{n+1}},\label{eta1}\\
  b&=&0,\label{b1}\\
  c&=&0.\label{c1}
\end{eqnarray}
We find $\eta<\epsilon$, and $b$ and $c$ satisfy the slow-roll conditions. Thus we should test whether $\epsilon$ and $\eta$ also satisfy the slow-roll conditions later. According to the former analysis, $u^2$ has the same amplitude of order as $|K(\phi)|$, and $|K(\phi)|$ is a decreasing function over time since the total energy density decreases during the inflation. Thus in terms of the model $K(\phi)=-K_{0}^2\phi^{2n}$, the inflaton field $\phi$ decreases with time during the slow-roll inflation phase. Then from the Eq.(\ref{epsilon1}), we find the slow-roll parameter $\epsilon$ decreases with time during the inflation, which should be satisfied during the inflation otherwise the inflation will last forever and destroy the thermal history of the Universe . When $\epsilon=1$, this marks inflation ends and exits to radiation dominated region smoothly. Then by Eq.(\ref{epsilon1}), we can calculate out the value of inflaton field when the inflation ends,
\begin{equation}\label{endfield}
  \phi_{e}^{n+1}=\frac{4\sqrt{3}nM_{p}}{K_{0}}.
\end{equation}
From Eqs.(\ref{efold}) and (\ref{endfield}), we can obtain the total number of e-folds of inflation ($\phi$ decreases with time thus $\sigma=-1$),
\begin{eqnarray}
  N&=&-\frac{1}{2\sqrt{3}M_{p}}\int_{\phi_{i}}^{\phi_e} K_0\phi^n d\phi\nonumber\\
  &=&\frac{2n}{n+1}\Bigg[\Bigg(\frac{\phi_i}{\phi_e}\Bigg)^{n+1}-1\Bigg]\nonumber\\
  &=&\frac{2n}{n+1}\Bigg(\frac{1}{\epsilon_{i}}-1\Bigg),
\end{eqnarray}
where $\epsilon_i$ is the initial value of $\epsilon$ when inflation begins. Usually, when the number of e-folds $N$ takes 60 or more bigger value, the inflationary paradigm can solve the problems of standard cosmology model. If we take $N=60$, when $n=1$, we get $\phi_i=7.75\phi_e$, $\epsilon_i=0.016$; $n=2$, we get $\phi_i=3.56\phi_e$, $\epsilon_i=0.022$; $n\gg1$, we get $\epsilon_i=0.032$. Thus if $\phi_i$ and $\phi_e$ satisfy the above relation, $\epsilon$ satisfies the slow-roll condition $\epsilon\ll1$ when the inflation begins, and the case can get enough number of e-folds of inflation.

However, there is a problem in this argument. In the former stability analysis, we find $r$ should remain smaller than the kinetic function $|K(\phi)|$. But when $\Gamma=\Gamma_{0}$,  $r$ can't remain smaller than $|K(\phi)|$ during the inflation which is inconsistent with the analysis in Sec.V since $r\propto\phi^{-2n}$ and $|K(\phi)|\propto\phi^{2n}$ by Eqs.(\ref{model1}) and (\ref{r}). Hence, the model of $\Gamma=\Gamma_0$ with a power-law form of $K(\phi)$ is not a workable model. If we assume that the dissipative coefficient has the form of $\Gamma\propto\phi^{m}$, we will avoid this problem. Next we will calculate the case of $\Gamma$ being a function of $\phi$ in order to obtain a self-consistent warm k-inflation model with a power-law form of $K(\phi)$.

\subsection{$\Gamma=\Gamma(\phi)$ case.}

We make the dissipative coefficient $\Gamma$ to be of the form $\Gamma=\Gamma(\phi)=\Gamma_m\phi^{m}$ where $\Gamma_m$ is a positive constant and $m$ is an even integer. Then the Hubble parameter is also given by Eq.(\ref{H}), and the dissipative rate $r$ and the radiation $\rho_{\gamma}$ are
\begin{eqnarray}
r&=&\frac{2\Gamma_m M_p\phi^{m-n}}{\sqrt{3}K_0|u|},\\
\rho_{\gamma}&=&\frac{\sqrt{3}\Gamma_m M_p\phi^{m-n}|u|}{2K_0},
\end{eqnarray}
respectively. Since $u\propto\phi^{n}$, $r$ is proportional to $\phi^{m-2n}$. If $m>4n$, the dissipative rate $r$ can remain smaller than the function $|K(\phi)|$ during the inflation. Moreover, the slow-roll parameters $\epsilon$, $\eta$ and $c$ are also given by Eqs.(\ref{epsilon1}), (\ref{eta1}) and (\ref{c1}), and $b$ is
\begin{equation}
  b=\frac{2\sqrt{3}mM_p}{K_0\phi^{n+1}}.
\end{equation}
When $m>4n$, we find $b\simeq\epsilon\simeq\eta$. Thus if $\epsilon$ satisfy the slow-roll conditions, $\eta$ and $b$ also meet them. Since $\epsilon$ is the same as that in the case of $\Gamma=\Gamma_0$, the final value of inflaton field $\phi_{e}$ is also same. Then we calculate the total number of e-folds of inflation,
\begin{equation}
  N=\frac{2n}{n+1}\Bigg(\frac{1}{\epsilon_{i}}-1\Bigg),
\end{equation}
which is also same as the former case. Thus it implies this case can also satisfy the slow-roll conditions and get enough number of e-folds of inflation. Then we give a kind of self-consistent warm k-inflation model with $K(\phi)=-K_{0}^2\phi^{2n}$ and $\Gamma=\Gamma_m\phi^m(m>4n)$ which can get enough number of e-folds of inflation and exit to the radiation dominated phase smoothly.

\section{\label{sec6}exponential form of $K(\phi)$}

In this section, we take the kinetic function $K(\phi)$ as to be of the exponential form:
\begin{equation}\label{exp}
  K(\phi)=-K_{1}^{2}e^{2\alpha\phi},
\end{equation}
where $K_1$ and $\alpha$ are both positive constants. This kind form of $K(\phi)$ has been also turned out that it can result in kinetically driven inflation in the ref.\cite{Mukhanov1999}. Next we also take $\Gamma$ as to be the forms of $\Gamma=\Gamma_0=constant$ and $\Gamma=\Gamma(\phi)$ respectively to give another kind of self-consistent warm k-inflation model.
\subsection{$\Gamma=\Gamma_{0}$ case.}
If $\Gamma=\Gamma_{0}$, we can calculate out the Hubble parameter and the dissipative rate:
\begin{eqnarray}
  H&=&\frac{K_1e^{\alpha\phi}|u|}{2\sqrt{3}M_p},\label{H2}\\
  r&=&\frac{2\Gamma_0M_p}{\sqrt{3}K_1e^{\alpha\phi}|u|},\label{r2}
\end{eqnarray}
respectively. The radiation density during the slow-roll inflation is given by
\begin{equation}
\rho_{\gamma}=\frac{\sqrt{3}\Gamma_0M_p|u|}{2K_1e^{\alpha\phi}}.
\end{equation}
Since $u^2$ has the same order of amplitude as $|K(\phi)|$,  we take $u$ as to be the form, $u\propto e^{\alpha\phi}$. Then by Eq.(\ref{r2}), we find $r\propto e^{-2\alpha\phi}$. Further, the slow-roll parameters can be given by
\begin{eqnarray}
\epsilon&=&\frac{4\sqrt{3}\alpha M_p}{K_1e^{\alpha\phi}}, \label{epsilon2}\\
\eta&=&\frac{4\sqrt{3}\alpha M_p}{K_1e^{\alpha\phi}}, \label{eta2}\\
b&=&0,\\
c&=&0.\label{c2}
\end{eqnarray}
We find $b$ and $c$ satisfy the slow-roll conditions and $\epsilon=\eta$. Then we begin to test whether $\epsilon$ and $\eta$ also satisfy the slow-roll conditions. On the basis of the former analysis, $|K(\phi)|$ is decreasing with time during the inflation. Thus in terms of the exponential form of $K(\phi)$ the inflaton field $\phi$ should decrease with time during the inflation. By Eq.(\ref{epsilon2}), we can know $\epsilon$ is a increasing function, so when $\epsilon=1$ implies the end of inflation. By Eq.(\ref{epsilon2}), we can calculate out the final value of infalton field,
\begin{equation}\label{endfield2}
\phi_e=\frac{1}{\alpha}\texttt{ln}\, \Bigg(\frac{4\sqrt{3}\alpha M_p}{K_1}\Bigg),
\end{equation}
According to the Eqs.(\ref{efold}) and (\ref{endfield2}), we can obtain the total number of e-folds,
\begin{eqnarray}\label{N}
N&=&-\frac{1}{2\sqrt{3}M_p}\int_{\phi_i}^{\phi_e} K_{1}e^{\alpha\phi} d\phi \nonumber\\
&=&2\bigg[e^{\alpha(\phi_i-\phi_e)}-1\bigg] \nonumber\\
&=&2\Bigg(\frac{1}{\epsilon_{i}}-1\Bigg).
\end{eqnarray}
Usually, when the number of e-folds $N$ takes $60$ or more bigger value, the inflationary paradigm can solve the problems of standard cosmology model. Then if we take $N=60$ we can deduce $\epsilon_i=0.032$ and $\phi_i=\phi_e-\frac{3.44}{\alpha}$ by Eq.(\ref{N}). Therefore, as long as the inflaton field takes proper initial value, $\epsilon$ and $\eta$ can satisfy the slow-roll conditions and get enough number of e-folds of inflation.

However, there is also a problem in this argument. When $\Gamma=\Gamma_0$, $r$ can't remain smaller than $|K(\phi)|$ during the inflation which is inconsistent with the analysis in Sec.\ref{sec5} since $r\propto e^{-2\alpha\phi}$ and $|K(\phi)|\propto e^{2\alpha\phi}$ by Eqs.(\ref{exp}) and (\ref{r2}). So the warm k-inflation model with $K(\phi)=-K_{1}^2e^{2\alpha\phi}$ and $\Gamma=\Gamma_0$ isn't a kind of self-consistent model. But if $\Gamma=\Gamma(\phi)$, we will avoid the above problem. Next we will try to study the case of $\Gamma=\Gamma(\phi)$.

\subsection{$\Gamma=\Gamma(\phi)$ case.}

When $\Gamma=\Gamma(\phi)=fe^{a\phi}$ where $f$ and $a$ are both positive constants. The Hubble parameter is also given by Eq.(\ref{H2}), and the dissipative rate and the radiation density are
\begin{eqnarray}
r&=&\frac{2fM_pe^{(a-\alpha)\phi}}{\sqrt{3}K_1|u|},\\
\rho_{\gamma}&=&\frac{\sqrt{3}fM_pe^{(a-\alpha)\phi}|u|}{2K_1},
\end{eqnarray}
respectively. Since $u\propto e^{2\alpha\phi}$, $r$ is proportional to $e^{(a-2\alpha)\phi}$. Thus if $a>4\alpha$, $r$ can remain smaller than the kinetic function $|K(\phi)|$ during the slow-roll inflation. Furthermore, $\epsilon$, $\eta$ and $c$ are given by Eqs.(\ref{epsilon2}), (\ref{eta2}) and (\ref{c2}), and $b$ is given by
\begin{equation}
b=\frac{2\sqrt{3}aM_p}{K_1e^{\alpha}}.
\end{equation}
When $a>4\alpha$, we find $b\simeq\epsilon=\eta$. Thus if $\epsilon$ satisfy the slow-roll conditions, $\eta$ and $b$ also meet. Since $\epsilon$ is the same as that in the case of $\Gamma=\Gamma_0$, the final value of inflaton field $\phi_{e}$
and the total number of e-folds of inflation are also same. Thus, the warm k-inflation model with $K(\phi)=-K_{1}^2 e^{2\alpha\phi}$ and $\Gamma=fe^{a\phi}$ ($a>4\alpha$) can also get proper number of e-folds of inflation and exit to the radiation dominated phase smoothly.

\section{\label{sec7}conclusions}

In this paper we firstly extend the k-inflation proposed by Mukhanov under the standard inflation theory to the warm inflation theory and give the dynamical equations of warm k-inflation. Then we also obtain the attractive fixed point $X_{f}$ which is similar to that of the k-inflation in addition to the correction of the dissipative rate $r$. Since the inflationary cosmology is often associated with a slow-roll solution, we rewrite the slow-roll parameters in the warm k-inflation which are very different from those in the potential driven models. The reliability of the slow-roll approximation demands that the slow-roll solution acts as an attractor for the dynamical system. We make a linear stability analysis to obtain the sufficient conditions for the reliability of the slow-roll solution. By the stability analysis, we yields the slow-roll conditions: $\epsilon\ll1$, $\vert \eta\vert \ll \frac{\mathcal{L}_X}{(\mathcal{L}_X+r)c_{s}^2}$, $ \vert b \vert \ll1$, $\vert c \vert <4$ and $\frac{rc_{s}^2}{\mathcal{L}_X}\ll1-2c_{s}^2$. The first two conditions mean the slow-roll solutions are quasi exponential solutions. The constraint on $\epsilon$ is same as that in the k-inflation under standard inflationary theory, and the constraint on $\vert \eta \vert$ implies that $K_{\phi}$ varies slowly during the slow-roll inflation.  $\vert b \vert \ll1$ means that $\Gamma$ is varying slowly together with the inflaton field.  The constraint on parameter $c$ is the same as that of the usual warm inflationary models. This implies that the dependence of the dissipative coefficient should be within the range of $\Gamma\propto(T^{-4}, T^4)$. The last condition provide some constrains on the kinetic function $|K(\phi)|$ of our model, thus we find $\frac{1}{2}(u^2+r)<|K(\phi)|\leq u^2$. Besides, $u^2$ should have the same order of the amplitude of $|K(\phi)|$ while the dissipative rate $r$ remains smaller than them during the inflation.

With the slow-roll conditions obtained, we study two cases of kinetic function $K(\phi)$ for the inflaton field: a power-law case ($K(\phi)=-K_{0}^2\phi^{2n\phi}$) and an exponential case ($K(\phi)=-K_{1}^2e^{2\alpha\phi}$). Both cases have the feature of kinetically driven inflation and the kinetic function $|K(\phi)|$ is decreasing during the inflation. Besides, in both cases we consider a constant dissipative coefficient and it being a function of the inflaton field $\phi$. Through the analysis of specific examples, we find $\Gamma=\Gamma_0=constant$ in both cases to not be a suitable choice since it can't ensure that the dissipative $r$ remains smaller than $|K(\phi)|$ during the slow-roll inflation. And we find $\epsilon$ is an increasing function during the inflation in warm k-inflation, which is consistent with the requirement of usual inflation. Otherwise, the inflation will last forever and destroy the thermal history of the Universe, unless some new mechanism ends the inflation. Finally, we obtain concrete reasonable models when the dissipative coefficient $\Gamma$ is a proper function of inflaon field $\phi$. There could be other consistent conditions for the warm k-inflation models, which deserve more research.

We leave to future calculate the investigation of general important issues: the choice of initial conditions and the computation of the perturbation spectra generated by this new kind of inflationary model.

\acknowledgments This work was supported by the National Natural Science Foundation of China (Grants No. 11575270, No. 11175019, No. 11235003 and No.11605100).

\end{document}